\documentclass[%
 reprint,
 amsmath,amssymb,
 aps
]{revtex4-2}

\usepackage{graphicx}
\usepackage{dcolumn}
\usepackage{bm}
\usepackage{hyperref}
\usepackage[mathlines]{lineno}

\newcommand{\ee}{e^+e^-}
\newcommand{\tautau}{\tau^+\tau^-}
\newcommand{\mumu}{\mu^+\mu^-}

\newcommand{\Htautau}{H\!\to\!\tautau}

\newcommand{\vmin}{v_{min}}
\newcommand{\sqrts}{\sqrt{s}}

\begin{document}

\preprint{}

\title{Higgs Boson Spookiness: Probing Quantum Nonlocality with Spacetime-Resolved \texorpdfstring{$\bm{H\!\to\!\tau^+\tau^-}$}{H -> tau+tau-} Decays}

\author{Lawrence Lee} 
\email{lawrence.lee.jr@cern.ch}
\affiliation{University of Tennessee, Knoxville, Tennessee, 37996}

\author{John Lawless} 
\affiliation{University of Tennessee, Knoxville, Tennessee, 37996}

\author{Caroline Riggall} 
\affiliation{University of Tennessee, Knoxville, Tennessee, 37996}

\date{\today}

\begin{abstract}
We demonstrate that a future precision $\ee$ Higgs factory would be able to perform a spacetime-resolved test of quantum nonlocality in Higgs boson decays.  In simulated $\ee\!\to\!ZH\!\to\!(\mumu)(\tautau)$ events at $\sqrts=240$~GeV, we reconstruct $\tau$ lepton decay vertices and measure spin correlations as a function of the spacetime interval between the two $\tau$ decays. Such a measurement would be able to test Bell-inequality-violating correlations for spacelike-separated decays, enabling direct exclusion of superluminal, finite-speed entanglement signaling theories. With 0.75~ab$^{-1}$ of integrated luminosity, entanglement signal propagation speeds below $\approx2c$ can be excluded at 95$\%$~CL. Signals establishing any spin correlation can be excluded for speeds below $\approx9c$. This constitutes the first proposed spacetime-resolved measurement of electroweak quantum entanglement at a particle collider and demonstrates a unique capability of future Higgs factories.
\end{abstract}

\maketitle


\section{Introduction}
\label{sec:intro}

One of the most striking features of quantum mechanics is its nonlocal nature. The crucial insights of Bell~\cite{Bell:1964kc} and subsequent experimental confirmations~\cite{Aspect:1981nv,Tittel:1997bi,new_test} showed us that quantum states represent behavior established outside of spacetime or via superluminal connection. The classic form of bipartite entanglement involves two particles with spin measurements that remain correlated regardless of the spacetime separation of non-commuting measurements.

In recent years, there has been a growing interest in fundamental quantum tests from the particle physics community. There has been much excitement in entanglement in top quark pairs~\cite{Afik:2020onf,Fabbrichesi:2021npl,Severi:2021cnj,Dong:2023xiw,Han:2023fci,White:2024nuc,ATLAS:2023fsd,CMS:2024pts}, $\tau$ pairs~\cite{Yang:2026uwu,Han:2025ewp,Altakach:2022ywa,Ehataht:2023zzt,Ma:2023yvd,Gabrielli:2026tnl}, and vector boson pairs~\cite{Aoude:2023hxv,Aguilar-Saavedra:2022mpg,Aguilar-Saavedra:2022wam,Gabrielli:2026tnl}, on top of a long history of entanglement in $\Lambda$, $B$-hadron, and Kaon pair production~\cite{Hao:2009kj,Pei:2025yvr,Faldt:2013gka,Afik:2025grr,Gisin:2000jn}. However, for as long as there have been quantum entanglement proposals for colliders, there have been objections to their validity~\cite{Abel:1992kz,Dreiner:1992gt,Abel:2025skj,Bechtle:2025ugc}. The objections correctly highlight that any angular correlation measurement from particle decays or scattering angles can be reproduced with a local hidden variable (LHV) theory, and therefore do not constitute a Bell test. We argue that this crucial loophole is made even larger for existing proposals and measurements by the fact that even timelike signaling could reproduce the results of such experiments.

In this work, we demonstrate that CHSH entanglement with explicitly reconstructed spacelike separation can be measured at future  collider experiments, with unique measurements involving the keystone of the Standard Model (SM), the Higgs boson $H$. As the only spin-0 (scalar) elementary particle in the SM, $H$ decays produce pairs of particles in a spin-singlet state. Every $H$ decay gives an interesting quantum state, but measuring these correlations is the central challenge.

In modern collider experiments, there is no direct way to measure the spin of directly measured particles. Instead, the parity-violating nature of the Weak force provides an opportunity. Particle decays via a $W$ current will maximally prefer couplings to left-handed particles and right-handed anti-particles. In the relativistic limit, these chirality eigenstates coincide with helicity eigenstates such that the decays encode parent spin state information in the kinematics of the decay products. By encoding information about the particle state in macroscopically measurable momenta, this decay might be considered a measurement event. As the outgoing decay products travel significant distances and interact with the environment, any entanglement will be lost as the state decoheres.

The past work described above explores angular correlation measurements that use this spin-to-kinematics encoding to probe entangled states. However, without resolving the spacetime separation between the decay events, the observed correlations can be interpreted as causally connected measurement events. While a test of quantum correlations, these tests are not direct tests of the \emph{nonlocal} nature of quantum mechanics as there is no explicit  spacelike separation required.

To address this weakness, we turn again to the $\tau$ lepton. The $\tau$

\begin{itemize}
    \item decays via a $W$ current such that information about its spin state becomes measurable in its decay kinematics;
    \item can be pair-produced in the Standard Model (SM) via heavy bosonic resonances, including the only fundamental spin-0 state, the Higgs;
    \item and has a large proper lifetime of about $0.29$~ps, resolveable at contemporary detectors, particularly when highly boosted, as in $H$ decays.
\end{itemize}

\noindent In this paper, we argue that this measurable lifetime is the key to directly probing the nonlocal nature of spin entanglement at colliders. Details of the Weak interaction, in its parity violation and its heavy mediator scale, work together to give us measurable spin and lifetime information, respectively, giving a perfect lab for studying spin entanglement at the highest energies.

We study $e^+ e^-$ collisions operating in a Higgs factory configuration with $\sqrt{s}=240$~GeV producing on-shell $ZH$ events. We focus on the case where the $Z$ boson decays to $\mu^+ \mu^-$, and the Higgs decays to $\tau^+ \tau^-$. The cleanly measured $Z$ system and collision energy allow for the full $H$ four-momentum to be measured.

The spin-0 Higgs then decays to two spin-$\frac{1}{2}$ $\tau$ leptons such that the $\tau\tau$ system is in a maximally-entangled singlet spin state. Since $m_H/2\gg m_\tau$, the $\tau$s are highly boosted and can gain lab-frame decay lengths $\mathcal{O}(1)$~mm and beyond. While traveling this small but measurable distance in vacuum, the quantum coherence would survive until (at least) the decay of the $\tau$.

Section~\ref{sec:taus} describes reconstruction of the collision events and the procedure to recover $\tau$ lepton properties. In Section~\ref{sec:spincorrelation}, we discuss the spin correlation measurement, while in Section~\ref{sec:results} the reconstructed decay vertices are used to probe the nonlocal nature of the system. Possible implications of such a measurement are considered in Section~\ref{sec:impact}.

The WHIZARD~\cite{Kilian:2007gr} event generator is used to simulate the $e^+e^-$ collision conditions to produce $ZH\rightarrow\mu^+\mu^-\tau^+\tau^-$ events at $\sqrt{s}=240$~GeV, which are then processed in Pythia8~\cite{Bierlich:2022pfr,Ilten:2013yed} to decay the $\tau$s, using its sophisticated $\tau$ decay feature to correctly model spin correlations in multi-$\tau$ events. 4462 events in which both $\tau$ leptons decay to $\pi^{\pm}\nu_\tau$ were generated, corresponding to about 0.75~ab$^{-1}$ of data.

\section{Complete $\tau\tau$ System Reconstruction}
\label{sec:taus}

Using the Weak decay of the $\tau$ as a spin measurement requires the measurable kinematic information in the decay to be usefully correlated with the underlying spin state. The measurability of the spin state is given by the spin-analyzing power of the decay. For simplicity, we restrict this analysis to the $\tau^\pm\rightarrow\pi^{\pm}\nu$ channel, whose two-body nature maximally preserves the $\tau^\pm$ spin information in the measurable $\pi^\pm$ kinematics, where the polarimeter vector $\mathbf{h}(\tau^{\pm} \to \nu_{\tau}\pi^{\pm}) \propto \mathbf{p}_{\pi^{{\pm}}}$ and the spin analyzing power of 100\% is reduced only by experimental resolutions on the $\pi^\pm$ track parameter measurements. Requiring this one decay mode reduces the acceptance of the analysis to only $BR(\tau\rightarrow\pi^\pm\nu)^2\approx1.2\%$~\cite{ParticleDataGroup:2024cfk}. An extension of this measurement to include decays through a $\rho$ resonance, yielding an additional neutral pion, could increase the acceptance to about $13\%$ of $H\rightarrow\tau\tau$ events.
However, detector effects from the additional photon measurements further impact the experimental analyzing power, and this extension is left for future work.

In an $e^+e^-$ Higgs Factory, $ZH$ production, knowledge of the initial state, and precise measurement of the $Z$ (in this study, decaying to $\mu\mu$) allow us to fully reconstruct the momentum of the Higgs boson. This information, along with the seemingly lossy reconstruction of hadronically decaying $\tau$s, creates an over-constrained system. Ref.~\cite{Jeans_2016} demonstrated that one could use the measurement of the hadron tracks and their impact parameters, the measured $Z$ recoil, and the known masses of these SM particles to fully reconstruct the location $\vec{x}_\pm$ of each $\tau^\pm$ decay. This position, combined with the $\tau^\pm$ momentum measurement additionally gives the \emph{time} of each decay $t_\pm$. The Higgs Factory $ZH(\tau\tau)$ topology uniquely allows for the direct measurement of the spacetime location of each $\tau^\pm$ decay~\cite{Jeans_2016}.

Throughout this study, parametrized experimental resolutions inspired by the International Large Detector (ILD) are used~\cite{ILDConceptGroup:2020sfq}. We model the ILD response by Gaussian smearing of track parameters: 
\begin{equation*}
    \sigma(1/p_T) = \sqrt{(2\!\times\!10^{-5})^2 + (10^{-3}/(p_T\!\sin\theta))^2}\text{ GeV}^{-1}
\end{equation*}
for the momentum including low-momentum multiple scattering contributions,
\begin{equation*}
    \sigma(d_0) = \sqrt{5^2 + (10/(p \sin^{3/2}\theta))^2}\text{ }\mu\text{m}
\end{equation*}
for the impact parameter, and $\sigma(\theta,\phi) = 0.1$~mrad for track angles~\cite{Kawada:2015wea,ILDConceptGroup:2020sfq}.

\begin{figure}[tbp]
     \centering
     \includegraphics[width=0.45\textwidth]{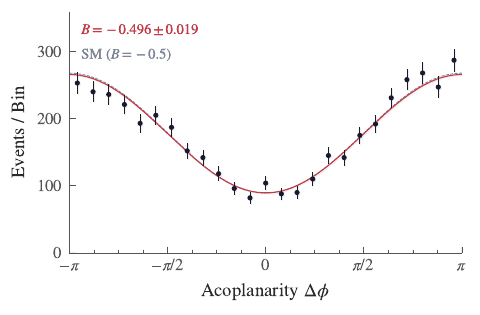}
     \caption{The angular correlation between the two polarimeter vectors is shown as measured in the Higgs rest frame. The SM expectation is shown in grey. Simulated points with realistic detector resolution are shown in black, and a cosine fit of this pseudo-data gives the red curve, in good agreement with the SM expectation.}
     \label{fig:angle}
\end{figure}

\section{Spin Entanglement Measurement in $H\rightarrow\tau\tau$ }
\label{sec:spincorrelation}

Measurements of the di-$\tau$ spin correlation can be performed by measuring the angle between the polarimeter vectors in the reconstructed Higgs rest frame. This correlation measurement is already a crucial one in Higgs physics to probe its CP nature~\cite{Jeans_2018,Grzadkowski:1995rx,Berge:2013jra,Bower:2002zx,Desch:2003mw,PhysRevD.92.096012}.

Figure~\ref{fig:angle} shows the expected distribution of events as a function of the acoplanarity $\Delta\phi$ of the two decay planes in the Higgs rest frame, as measured using the two polarimeter vectors. The SM (\emph{i.e. quantum}) expectation is a purely sinusoidal distribution that can be fit to the form $A(1+B\cos\Delta\phi)$ where $A$ and $B$ are fitted constants. The SM expectation is that $B=-0.5$, and no correlation would correspond to $B=0$~\footnote{BSM contributions to the CP nature of the Higgs would result in a phase shift of the cosine. Such a phase shift is not considered as a free parameter in the fit performed here.}.

The spin correlation matrix can be constructed to evaluate the degree of entanglement in the system; in this study, we use the Horodecki parameter $m_{12}$, the sum of the two largest eigenvalues of the matrix $CC^T$, where $C$ is the spin correlation matrix~\cite{Horodecki:1995nsk}. The CHSH inequality~\cite{Clauser:1969ny} corresponds to $m_{12}<1$, and violations of this inequality suggest nonlocal entanglement given spacelike-separated spin measurements. The SM value $m_{12}=2$ violates this inequality.

\section{Probing nonlocality in $H\rightarrow\tau\tau$}
\label{sec:results}

Without explicitly requiring it, the experimenter cannot guarantee that the $\tau$ decays are spacelike-separated. If timelike-separated, decay events can in principle be causally connected and no such measurement can probe quantum nonlocality. To address this loophole, we combine the methods of Sections~\ref{sec:taus} and \ref{sec:spincorrelation} to produce a spacetime-resolved measurement of the entangled spin states.

Since we are able to fully reconstruct the spacetime location of each $\tau$ decay in this $ZH$ topology, we can explicitly measure if the decays are spacelike-separated. From the reconstructed decay vertices $(t_\pm, \vec{x}_\pm)$ of the two taus, we compute the spatial separation $\Delta r = |\vec{x}_+ - \vec{x}_-|$ and the time difference $\Delta t = |t_+ - t_-|$ in the lab frame. For spacelike-separated events, no subluminal signal could connect the two decays. The minimum signal speed for causal connection is $\vmin = \Delta r/(c\,\Delta t)$. When $\vmin>c$, the decay events are spacelike-separated. A small timelike population (${\sim}5\%$) arises when one tau decays promptly while the other survives for several decay lengths. These events provide an internal consistency check that the same entanglement is observed even when a subluminal causal connection is allowed.

The momentum resolution has negligible impact on the spin correlation measurement;  the dominant effect is the impact parameter resolution. The typical $\tau$ impact parameter $d \sim L\!\sin\alpha \approx 2.4\;\mu$m is comparable to $\sigma(d_0)\approx 5\;\mu$m, yielding $d/\sigma(d_0)\approx 0.5$. This effect degrades the decay-length resolution significantly, broadening the measured spacetime interval distribution.

\begin{figure}[tbp]
     \centering
     \includegraphics[width=0.40\textwidth]{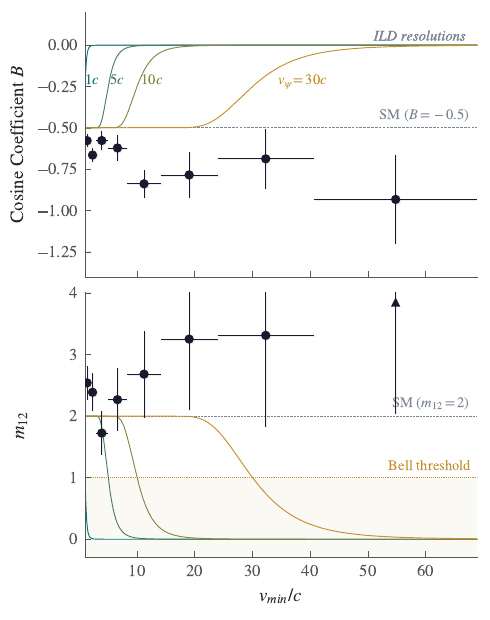}
     \caption{The fitted cosine amplitude $B$ (top) and Horodecki parameter $m_{12}$ (bottom) are shown as a function of the speed needed to connect measurement events $v_{min}$. The SM value of $m_{12}=2$, $B=-0.5$ and the Bell threshold of $m_{12}=1$ are shown as horizontal lines, simulated events with $\mathcal{L}=0.75$~ab$^{-1}$ are shown as markers, and example superluminal, finite $v_\psi$ quantum signaling hypotheses are shown assuming ILD-like detector resolutions.}
     \label{fig:speed}
\end{figure}

\begin{figure}[tbp]
     \centering
     \includegraphics[width=0.45\textwidth]{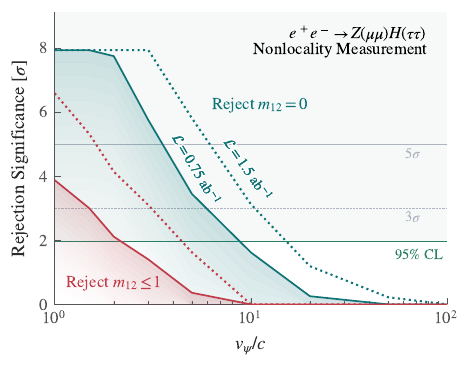}
     \caption{With an integrated luminosity of 0.75~ab$^{-1}$ (1.5~ab$^{-1}$), the potential rejection significance for this measurement for the $m_{12}=0$ and $m_{12}=1$ hypotheses is shown as a function of $v_\psi$ as solid (dotted) lines. Bell-threshold-level entanglement (red) can be rejected at 95\%~CL up to $v_\psi\approx2c$ ($5c$) and complete noncorrelation (blue) could be rejected up to $v_\psi\approx9c$ ($16c$).}
     \label{fig:signif}
\end{figure}

Figure~\ref{fig:speed} shows both the fitted cosine amplitude $B$ and $m_{12}$ for this 0.75~ab$^{-1}$ of simulated data as a function of $v_{min}$, along with expectations from the SM and example hypotheses of a superluminal, but finite, quantum signaling speed $v_\psi$ in the lab frame. If quantum influences did propagate (super-)causally at some $v_\psi$, we expect a ``turn-off'' effect for $v_{min}$ values exceeding $v_\psi$, as no quantum signal could travel fast enough to give rise to the measured correlation. Extreme limits of $\mathcal{O}(10^4c)$ in the lab frame have been set on $v_{\psi}$ in optical systems~\cite{Scarani:2000tjz,Gisin2008}, but these measurements have been at low energies. Here, we have an opportunity to measure this effect at the high-energy, electroweak scale of $\mathcal{O}(100)$~GeV.

In the simulated data sample using example experimental resolutions from the ILD design, a wide range of $v_{min}$ can be probed, to roughly $50c$, with SM expectations showing a constant correlation with $B=-0.5$ and $m_{12}=2$ across all values of $v_{min}$. In Figure~\ref{fig:signif}, we show the projected rejection sensitivity for the $m_{12}\le1$ (locality limit) and $m_{12}=0$ (no spin correlation) scenarios as a function of the hypothesized lab-frame $v_\psi$, assuming the SM central value of $m_{12}=2$ and statistical uncertainties given integrated luminosities of 0.75~ab$^{-1}$ and 1.5~ab$^{-1}$. Scenarios where all correlation disappears when the second decay is out of reach of a signal at $v_\psi$ in the lab frame would be rejected up to a speed of $v_\psi\approx9c$ and $16c$ for the two luminosity scenarios, respectively. The stronger statement of rejecting correlations at the Bell threshold can be rejected up to $v_\psi\approx2c$ and $5c$, respectively.

\section{Discussion}
\label{sec:impact}

This analysis would demonstrate the first
spacetime-resolved measurement of quantum entanglement at a particle
collider.  Unlike previous collider measurements that establish
entanglement by integrating over all events~\cite{ATLAS:2023fsd,CMS:2024pts},
the vertex reconstruction method of Ref.~\cite{Jeans_2016} allows direct access to the causal structure of each event pair.  This capability is unique to $\ee$
Higgs factories, where the known beam energy, clean environment, and
excellent vertex resolution combine to make the measurement feasible.
At the LHC, the unknown partonic center-of-mass energy and the large
backgrounds preclude equivalent vertex reconstruction in $\Htautau$~\footnote{Three-prong $\tau$ decays could allow for explicit vertex reconstruction without the over-constrained kinematic system used here. At the LHC, the beamspot resolution is however too large to allow for significant sensitivity.}.

Several extensions can substantially improve
the sensitivity:
(i)~including $Z\to e^+e^-$ doubles the leptonic-$Z$ sample;
(ii)~$Z\to q\bar{q}$ with hadronic $Z$ tagging increases the sample
by an order of magnitude, at the cost of modest Higgs-recoil
resolution;
(iii)~the $\tau\to\rho\nu$ channel ($\mathrm{BR}=25.5\%$) provides
additional events with reduced but nonzero analyzing
power~\cite{Tsai:1971vv};
(iv)~beam polarization can enhance the $ZH$ cross section by up to $50\%$~\cite{ILDConceptGroup:2020sfq}.

Because of the lack of an external choice of measurement direction, a crucial caveat to this proposal is that, to some degree, one must assume that there is no LHV encoding the eventual decay plane~\cite{Abel:1992kz,Dreiner:1992gt,Abel:2025skj,Bechtle:2025ugc}. This flaw is shared with all other proposed spin entanglement measurements at colliders, but resolving the spacetime separation of the decay events is a crucial step towards closing a related loophole. Resolving this separation would require that such hidden information be carried coherently by each $\tau$, instead of the decay kinematics being correlated locally between the two. In the least favorable interpretation, this test could accommodate LHVs, but purely angular tests can be explained with no ``spooky action at a \emph{distance}'' whatsoever.

This distinction exposes a hierarchy of loopholes that is absent in traditional Bell tests but relevant for any collider measurement. For analyses without explicit spacetime separation, an entirely local explanation is available that is strictly weaker than LHV: a subluminal signal ($v_\psi \leq c$) propagating from the first decay vertex to the second could in principle dictate the observed correlations through ordinary local causation, with no pre-shared hidden variables and no conspiracy at the production point.  This is a more fundamental loophole that requires no exotic assumptions. Bell's theorem does not apply, as it fundamentally assumes spacelike separation of measurement events. 

In contrast, the explicitly measured spacelike separation presented here removes this possibility. No quantum signal at $v_\psi \leq c$ can bridge a spacelike gap, so the only surviving classical explanation is LHV.  Superluminal, finite-speed quantum signaling is experimentally testable at these energy scales by observing correlations disappear at some separation. The observed Bell violation in this sample therefore constitutes a test of nonlocality in the strict sense that Bell's theorem requires. 

A direct probe of nonlocality in Higgs decays would, for the first time, probe this quantum effect at energies as high as the electroweak scale. Past measurements occurred at much lower energy densities, usually at the atomic scale. In the study of (local or nonlocal) hidden variables in quantum mechanics, David Bohm suggested that the quantum effects may only be a low energy approximation of a deeper theory that would emerge from studying quantum foundations at higher energy scales~\cite{bohmI}.  More crucially, tests of these nonlocal effects, in addition to being at lower energy, have been with quantum entanglement established via electromagnetic means. The test proposed here would be the first direct probe of the electroweak sector and its adherence to quantum nonlocality. As the Weak interaction has provided its share of surprises, foundational tests must be independently performed in this sector.

\section{Conclusion}
\label{sec:conclusion}

We have shown that spacetime-resolved entanglement measurements in $\Htautau$ are feasible at a future Higgs factory and provide a direct, model-independent test of quantum nonlocality with massive fermions.  The measurement constrains causal theories with superluminal, finite signal speed, probes the causal structure of quantum
correlations in a high-energy relativistic regime, and demonstrates a unique
physics capability of future Higgs factories. Similar studies should be done for $Z\rightarrow\tau^+\tau^-$ spin correlations for entanglement study opportunities at the proposed tera-$Z$ FCC-ee~\cite{Blondel:2018mad}.

\acknowledgments

We acknowledge the invaluable help of Daniel Jeans in discussing his $\tau\tau$ reconstruction method and for providing the Monte Carlo sample used in this study. We thank many people for useful discussions on this topic, useful comments on this work, and for their general guidance and insight: 
Karri Folan DiPetrillo, Innes Bigaran, Dorival Gonçalves, Matthew Low, Michael Peskin, Graham Wilson, Jane Cummings, Zbigniew Was, Max Scherzer, Alex Tuna, Zach Marshall, Sarah Demers, Witek Skiba, and many others. This work has been supported by the Department of Energy, Office of Science, under Grant No. DE-SC0020267 and the National Science Foundation, under Award No. 2235028. This work was enabled by a residency at Fermilab made possible by the LHC Physics Center and the Universities Research Association.

%

\bibliographystyle{unsrt} 
\bibliography{tauentanglement}

\end{document}